\documentclass[reprint,
nofootinbib,
 amsmath,amssymb,
 aps,
]{revtex4-2}

\usepackage[utf8]{inputenc} 
\usepackage[T1]{fontenc}    
\usepackage{hyperref}       
\usepackage{url}            
\usepackage{booktabs}       
\usepackage{amsfonts}       
\usepackage{nicefrac}       
\usepackage{microtype}      
\usepackage{graphicx}
\usepackage{xcolor, etoolbox}
\usepackage{amsmath,amsfonts,amsthm,mathtools} 
\usepackage{ragged2e}
\usepackage{braket}      
\usepackage[english]{babel} 
\graphicspath{{./figures/}} 
\usepackage{enumitem}
\usepackage{xspace}
\usepackage{dsfont}
\usepackage{bm}
\usepackage{bbm}
\usepackage{dcolumn}

\usepackage[colorinlistoftodos]{todonotes}
\usepackage{mathrsfs}

\newcommand{\rh}{\ensuremath{\rho}\xspace}
\newcommand{\sig}{\ensuremath{\sigma}\xspace}
\newcommand{\tr}{\text{Tr}}

\hypersetup{
    colorlinks,
    linkcolor=blue,
    citecolor=blue,
    urlcolor=blue
}

\begin{document}

\preprint{APS/123-QED}

\title{Second law of thermodynamics for relativistic fluids formulated with relative entropy}

\author{Neil Dowling}
 \email{dowling@thphys.uni-heidelberg.de}
\author{Stefan Floerchinger}
 \email{stefan.floerchinger@thphys.uni-heidelberg.de}
\author{Tobias Haas}
 \email{t.haas@thphys.uni-heidelberg.de}
 \affiliation{Institut f\"{u}r Theoretische Physik, Universit\"{a}t Heidelberg, \\ Philosophenweg 16, 69120 Heidelberg, Germany}


\begin{abstract}
The second law of thermodynamics is discussed and reformulated from a quantum information theoretic perspective for open quantum systems using relative entropy. Specifically, the relative entropy of a quantum state with respect to equilibrium states is considered and its monotonicity property with respect to an open quantum system evolution is used to obtain second law-like inequalities. We discuss this first for generic quantum systems in contact with a thermal bath and subsequently turn to a formulation suitable for the description of local dynamics in a relativistic quantum field theory. A local version of the second law similar to the one used in relativistic fluid dynamics can be formulated with relative entropy or even relative entanglement entropy in a space-time region bounded by two light cones. We also give an outlook towards isolated quantum field theories and discuss the role of entanglement for relativistic fluid dynamics.
\end{abstract}

\maketitle


\section{\label{sec:level1}Introduction\protect}

In recent years entanglement entropy has developed into a key concept in areas of quantum field theory (QFT) such as black hole physics \cite{Bombelli1986, Srednicki1993, Callan1994, Wall2011, Casini2008}, holography \cite{Ryu2006a,Ryu2006b, Casini2011} and high energy physics \cite{Kharzeev2017,Shuryak2017,Berges2018a,Berges2018b,Kovner2018,Armesto2019,Tu2019} (for general aspects and methods see refs. \cite{Casini2009, Calabrese2004, Witten2018}). It could play a role to better understand non-equilibrium dynamics of quantum fields and the emergence of relativistic fluid dynamics. An interesting hypothesis is that local dissipation in such fluids might be understood as the generation of entanglement.

Let us start with the density operator $\rho$ of a quantum system that can be split into two parts, $A$ and $B$. With the reduced density operator $\rh_A = \text{Tr}_{B} \{ \rh \}$ for the subsystem $A$, the global von Neumann entropy and entanglement entropy are defined to be, respectively \cite{vonNeumann1955,Nielsen2010}, 
\begin{equation} \label{eq:VNentropy}
    S(\rh) = - \text{Tr} \{ \rh \ln \rh \}, \ \ \ \ \ \ \ S_A(\rh) = - \text{Tr}\{ \rh_A \ln \rh_A \}.
\end{equation}

Entanglement entropy of a spatial region is ultraviolet (UV) divergent in a relativistic QFT according to an area law \cite{Casini2009}. The leading divergence is proportional to $\epsilon^{-(d-2)}$, where $d$ is the number of space-time dimensions and $\epsilon$ is a small length with $1/\epsilon$ acting as a UV momentum cutoff. These divergences depend on the geometry of the region but not on the state. This poses a fundamental problem in understanding the role entanglement plays within dynamical evolution in non-equilibrium QFT. In particular, one cannot easily formulate a local variant of the second law of thermodynamics, as it is phenomenologically used for example in relativistic fluid dynamics, based on the entanglement entropy $S_A$ of a subregion.

A possible solution to this problem could be to work instead with quantum relative entropy (the quantum version of the Kullback–Leibler divergence \cite{Kullback1951, Kullback1959}) which, given two density operators \rh and \sig, is defined as \cite{Umegaki1962}
\begin{equation}
    S(\rh \| \sig) =  \text{Tr} \{\rh \, (\ln \rh - \ln \sig) \}.
    \label{eq:defRelEntropy}
\end{equation}
In many cases the first argument \rh can be thought of as the actual system of interest, whereas the second argument \sig is some model system to compare with. Then the relative entropy quantifies the uncertainty deficit about \rh based on the false guess \sig. It gives a non-negative value, vanishes if and only if the density operators are equal and is finite given the support condition $\text{supp}(\rh) \subseteq \text{supp}(\sig)$. If this condition is violated the value can be set to $+\infty$ \cite{Nielsen2010,Cover2006,Vedral2002}. These properties qualify relative entropy to be a so-called {\it divergence}, but it is not a distance measure (metric) as it is not symmetric and does not satisfy the triangle inequality. 

Even for classical systems there are interesting and valid reasons for the use of relative entropy in some areas where entropy is currently being used. In contrast to Shannon entropy, there is a well defined continuous limit for relative entropy and a change of coordinates does not change its value as it does for differential entropy. For a relativistic quantum field theory, the relative entropy of two reduced density matrices for spatial subregions can be defined rigorously in terms of modular theory \cite{Araki1977}. A reformulation of the maximum entropy principle in the context of statistical physics in thermal equilibrium based on relative entropy is given in ref.\ \cite{Haas2020}. Furthermore, the use of relative entropy in QFT is discussed in refs. \cite{Casini2008,Witten2018,Lashkari2014,Lashkari2016,Arias2017,Ruggiero2017,Araki1977}.

In the current work relative entropy will be most useful to us in investigating stochastic evolution for open quantum systems. When investigating the second law for some arbitrary quantum state \rh, we need to choose some model \sig to compare it to. A suitable choice for us will be thermal equilibrium states. For example, when $\rho$ describes an open system in contact with a heat bath, it is convenient to choose $\sigma$ to be the density matrix of the canonical ensemble with inverse temperature $\beta=1/T$. This is not only the model with the highest entropy for the given physical situation, but as a density matrix also has a broad support so that the relative entropy $S(\rho \| \sigma)$ is well defined. It is straight forward to rewrite this relative entropy with respect to the canonical state as
\begin{align}
    S(\rh \| \sig) = - S(\rh) + S(\sig) + \beta \left[E(\rh) - E(\sig) \right],
    \label{eq:Relative_Entropy_Thermal}
\end{align}
where $E(\rh)=\tr \{ \rh H \}$ is the energy expectation value of the state described by \rh.

The relative entropy of two reduced density matrices $\rho_A = \text{Tr}_{B} \{ \rh \} $ and $\sigma_B = \text{Tr}_{B} \{ \sigma \}$ is also known as {\it relative entanglement entropy}, $S_A(\rh \| \sig) = S(\rho_A \| \sigma_A)$. In contrast to entanglement entropy, relative entanglement entropy does not show UV divergences and is expected to be generically finite. Intuitively speaking, the divergent terms, which are independent of the specific state of the quantum field theory, cancel out. For this reason we believe that relative entanglement entropy is well suited for investigating the dynamics of entanglement and non-equilibrium evolution in a quantum field theory.

Entanglement generation is assumed to be a driving mechanism behind thermalization and second law-like behavior of macroscopic quantum systems \cite{Popescu2006}. In terms of total entropy the thermalization of an isolated system is difficult to describe: the time evolution of an isolated system is a unitary map according to the von Neumann equation. Thus the total entropy remains constant over time since the von Neumann entropy of any state is invariant under a unitary transformation, 
\begin{equation}
    S(U \rh U^{\dagger}) = S(\rh).
\end{equation}
This implies that a perfectly isolated quantum system in this sense actually does not thermalize. However, one may investigate instead a subsystem of an isolated system and this may evolve non-unitarily due to the exchange of quantum information with its surroundings. The ``surroundings'' could here either by another sector of the theory, or it could be a neighboring spatial region. Such a subsystem is then referred to as an open quantum system.
 
Mathematically, the evolution of an open quantum system is described by some {\it completely positive trace-preserving map} $\mathcal{N}$ (CPTP map or quantum channel). In full generality a CPTP map describes communication of quantum information between quantum states, encompassing any map of a density matrix to some other. Specifically, the Kraus theorem \cite{Kraus1983,Ozawa1984} allows one to decompose these maps as 
\begin{equation}
    \mathcal{N}: \quad \rh \rightarrow \mathcal{N}(\rh) = \sum_{\alpha} A_\alpha \rh A_\alpha^{\dagger},
\end{equation}
where \(A_\alpha\) are Kraus operators fulfilling the condition \(\sum_{\alpha} A_\alpha^{\dagger} A_\alpha  = \mathbbm{1}\). CPTP maps account for many possible quantum transformations, including for example unitary time evolution and partial tracing over a subsystem.

If a CPTP map is applied to a quantum state, its von Neumann entropy may increase as well as decrease. An example of an entropy decreasing process would be the loss of heat of a system due to cooling\footnote{A more explicit example can be seen in a generalized measurement as follows. Consider a maximally mixed two-state quantum system $\rh$ under the action of the non orthogonal measurement operators $M_1=\ket{0}\bra{0}$ and $M_2 = \ket{0}\bra{1}$. Then the state after measurement without recording the result, $\rh_M = M_1\rh M^{\dagger}_1 + M_2 \rh M^{\dagger}_2$, has a smaller entropy $S(\rh_M)< S(\rh)$ \cite{Nielsen2010}.}. CPTP maps are thus more general than the stochastic processes that imply the second law of thermodynamics. There is nonetheless a useful subclass of maps; a unital CPTP map $\mathcal{N}_I$, for which $\mathcal{N}_I (\mathbbm{1})=\mathbbm{1}$ holds, never decreases entropy,
\begin{align}
    S(\mathcal{N}_I (\rh)) \ge S(\rh). \label{eq:UnitalEntropy}
\end{align}
CPTP maps are central to the monotonicity of relative entropy, the quantum information theorem we will utilize to formulate a second law. It states that no quantum channel can increase distinguishability between states \cite{Nielsen2010,Vedral2002,Cover2006,Fuchs1996},\footnote{Recently this inequality was proven for the more general case of positive trace-preserving maps \cite{Reeb2017}. Furthermore, a strengthened version, which exhibits a remainder term from a rotated Petz recovery map, was established in ref. \cite{Berta2015}.}
\begin{equation}
    S(\mathcal{N}(\rh)\|\mathcal{N}(\sig)) - S(\rh\|\sig) \leq 0 . \label{monotonicity}
\end{equation}
Below we will employ this property to make statements about (local) thermalization or second law-like behavior by investigating a state $\rh$ approaching an equilibrium state $\sig$ measured in terms of quantum relative entropy or relative entanglement entropy. In the present work we will use a subclass of CPTP maps which keep the reference state \sig as a steady state invariant, $\mathcal{N}(\sigma) = \sigma$. This may be seen as an alternate definition describing stochastic evolution. For ordinary thermodynamics, second law-like inequalities from relative entropy are discussed in \cite{Sagawa2012}.

We will further develop a local formulation of a second law in the context of a quantum field theory in open exchange of quantum information with a bath fluid. This is also motivated by the aim to understand the relation between quantum field theory and relativistic fluid dynamics. While phenomenologically relativistic fluid dynamics seems to be a good approximation to quantum field dynamics, for example for the quark-gluon plasma created in heavy ion collisions \cite{Heinz:2013th, Busza:2018rrf, Teaney2009}, the detailed relation is yet to be properly understood. Usually in the context of fluid dynamics, a local second law is postulated and stated in terms of an entropy four-current density (e.\ g.\ see refs. \cite{Israel1979, Landau1987, Kovtun2012}),
\begin{align} \label{eq:SecondLawCurrent}
    \nabla_{\mu} s^{\mu}(x) \ge 0.
\end{align}
One problem here is that an entropy current $s^\mu(x)$ is difficult to define outside of global thermal equilibrium and if one aims to work with entanglement entropy one faces the same problems of UV divergence we described previously. As an alternative, we here propose a formulation in terms of {\it relative entropy} of a true state $\rh$ with respect to some form of equilibrium reference state $\sig$, with a second law-type inequality that essentially follows from the monotonicity property (see section \ref{sec:LocalSecondLaw}). This makes a step towards understanding quantum field theory in the fluid dynamic regime from a quantum information theoretic perspective.

The paper is structured as follows: firstly in section \ref{sec:Relativisticfluiddynamics} we recall some elements of relativistic fluid dynamics. In section \ref{sec:thermosecondlaw} we will outline the connection between the general thermodynamic second law and relative entropy, and then in section \ref{sec:LocalSecondLaw} we will develop the local form of the second law in the relativistic fluid dynamic regime for a causally complete space-time region. Finally, we will draw some conclusions in section \ref{sec:Conclusions}.

\begin{figure}[t!]	
        \includegraphics[width=0.48 \textwidth]{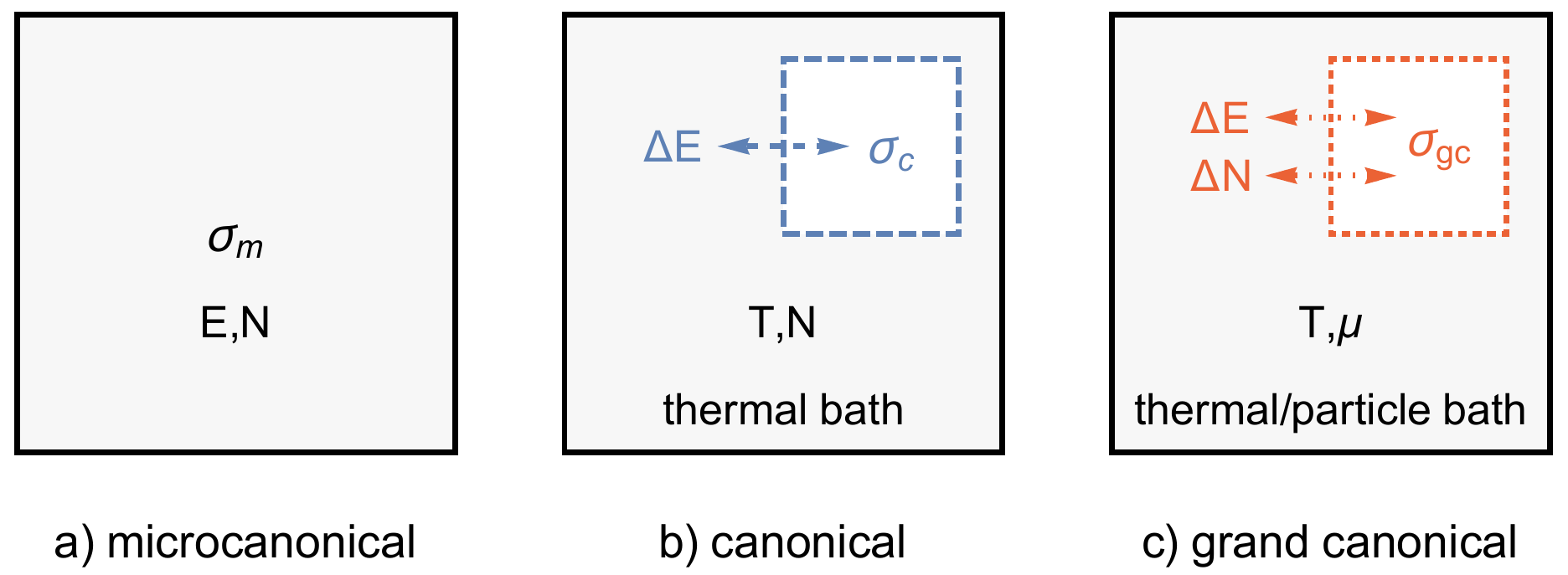}
        \caption{The three different statistical ensembles: (a) a system closed to heat and particle exchanged, (b) a system coupled to a heat bath allowing for energy exchange and (c) a system coupled to a heat and particle bath permitting energy and particle exchange.} \label{fig:Ensembles}
\end{figure}

\paragraph*{Notation.}
In this paper we adopt natural units, with $\hbar=c=k_B=1 $ and work with the Minkowski space metric signature $(-,+,+,+)$. Hats on operators are dropped. Expectation values are expressed with the relevant density operator as a parameter so that $\mathcal{O} (\rh)=\langle \mathcal{O} \rangle_{\rh} = \text{Tr}\{ \rh  \mathcal{O}\}$.

\section{Relativistic fluid dynamics}
\label{sec:Relativisticfluiddynamics}
Relativistic fluid dynamics can be seen as an effective description of (quantum) field theoretic degrees of freedom in out-of-equilibrium situations. It uses the concept of a {\it local} thermal equilibrium and an expansion around this, however. In the following we shall recall the construction with a perspective from quantum information theory.

Usually one starts from covariant conservation laws, such as for energy and momentum \cite{Israel1979,Landau1987,Kovtun2012}. This is a consequence of diffeomorphism symmetry if the theory is formulated in general coordinates with Riemannian metric $g_{\mu \nu}(x)$,
\begin{equation}
    \nabla_\mu T^{\mu \nu}(x) = 0.
    \label{eq:EnergyMomentumConservation}
\end{equation}
In addition the theory may exhibit a U$(1)$ symmetry leading to a covariantly conserved particle number current,
\begin{equation}
    \nabla_\mu N^{\mu}(x) = 0.
    \label{eq:ParticleNumberConservation}
\end{equation}

Furthermore, one also introduces an entropy current $s^{\mu}(x)$. In a phenomenological approach it is postulated to be governed by a local form of the second law
\begin{equation}
    \nabla_\mu s^{\mu}(x) \ge 0,
    \label{eq:LocalSecondLaw}
\end{equation}
where equality is reached in thermal equilibrium. Unlike the two former equations the local second law does not follow from symmetry considerations and needs a more careful justification. Moreover, it is not clear whether a local entropy current is well-defined in out-of-equilibrium situations or how precisely it can be defined from a microscopic quantum field theory \cite{Floerchinger2016}. In the following we investigate for what states an entropy current can be defined and also formulate an alternative to \eqref{eq:LocalSecondLaw} using relative entropy.

One should note that the above equations could be supplemented by additional conservation laws or equations for additional order parameters.  

With the conservation relations \eqref{eq:EnergyMomentumConservation} and \eqref{eq:ParticleNumberConservation} as well as eq.\ \eqref{eq:LocalSecondLaw}, one can discuss relativistic thermodynamics. In thermal equilibrium one can assume the entropy current to be a function of the conserved energy-momentum tensor and particle current $s^\mu(T^{\lambda\nu}, N^\sigma)$, and write 
\begin{equation}
    \nabla_\mu s^{\mu} = \frac{\partial s^{\mu}}{\partial T^{\lambda \nu}} \nabla_\mu T^{\lambda \nu} + \frac{\partial s^{\mu}}{\partial N^{\sigma}} \nabla_\mu N^{\sigma}.
\end{equation}
Because \eqref{eq:LocalSecondLaw} should reduce to an equality in thermal equilibrium as a consequence of the two covariant conservation laws \eqref{eq:EnergyMomentumConservation} and \eqref{eq:ParticleNumberConservation}, one should have
\begin{equation}
    \frac{\partial s^{\mu}}{\partial T^{\lambda \nu}} = - \beta_\nu \, \delta^{\mu}_{\lambda}, \hspace{1cm} \frac{\partial s^{\mu}}{\partial N^{\sigma}}= - \alpha \, \delta^{\mu}_{\sigma}.
\end{equation}
Here $\beta^\nu$ is a vector field and $\alpha$ is a scalar field, which together serve as parametrisation for the covariantly conserved fields in thermal equilibrium. These two fields correspond to the ratio of fluid velocity $u^\nu$ and temperature $T$ as well as chemical potential $\mu$ and temperature, respectively,
\begin{equation}
    \beta^\nu = \frac{u^\nu}{T}, \hspace{1cm} \alpha  = \frac{\mu}{T}.
    \label{eq:DefThermodynamicFields}
\end{equation}

Because $\nabla_\mu s^\mu$ must not only vanish in equilibrium but also be stationary, one finds for its differential
\begin{equation}
\nabla_\mu d s^\mu = - \nabla_\mu \beta_\nu d T^{\mu\nu} - \partial_\mu \alpha \, dN^\mu = 0,
\end{equation}
which leads to the condition that $\beta^\nu$ must be a Killing vector field and $\alpha$ a constant,
\begin{equation} 
    \nabla_\mu \beta _\nu + \nabla_\nu \beta _\mu = 0, \quad\quad\quad \partial_\mu \alpha = 0.
    \label{eq:KillingEquation}
\end{equation}
While $\beta^\nu$ and $\alpha$ are well defined in thermal equilibrium, there is some freedom in their definition outside of equilibrium. For example, the fluid velocity $u^\mu$ could be related to energy flow (the Landau frame definition), to the particle number flow $N^\mu$ (the Eckart frame definition) or be defined otherwise.

One of our main goals in the following will be to understand better how the local form of the second law of thermodynamics, eq.\  \eqref{eq:LocalSecondLaw}, or a variant of it, can arise from quantum field theory. We will argue that a formulation based on relative entropy has advantages in this context.

As a preparation, we discuss now first a global formulation of the second law based on relative entropy for a generic (open) quantum system.

\section{Thermodynamics: A General Second Law from Relative Entropy} \label{sec:thermosecondlaw}

In this section we will consider a generic quantum system coupled to an external bath with which it may exchange quantum information. In addition, there may also be an exchange of energy and / or particle number, but that does not have to be the case. 
We want to discuss how one can obtain a second law Clausius inequality from the monotonicity of relative entropy, and thus show the equivalence of the former with a relation written solely in terms of relative entropy. Many elements of this have already been investigated in ref.\ \cite{Sagawa2012} but we recall them here in order to prepare for a subsequent extension to quantum field theory.

The second law will be derived through a comparison of an arbitrary state $\rh$ with a suitable statistical ensemble or model state $\sig$ given the physical situation (with or without exchange of energy or particle number with the bath, for an overview see figure \ref{fig:Ensembles}). The corresponding equilibrium density operators $\sig$ follow from maximizing von Neumann entropy $S(\sig)$ under the appropriate constraints \cite{Jaynes1957,Jaynes19572,Jaynes1963,Jaynes1968,Landau1980}. Alternatively this can be done from minimizing an \textit{expected} relative entropy, as shown recently in ref.\ \cite{Haas2020}. The intensive thermodynamic quantities like temperature $T$ and chemical potential $\mu$ are chosen such that they agree with those induced by the surroundings. 
 
An open quantum system in a fixed volume $V$ evolves generically according to some CPTP map $\mathcal{N}$. Furthermore, we will take $\mathcal{N}$ to be within a subclass of CPTP maps that admits the relevant equilibrium state \sig as a steady state, $\mathcal{N}(\sig)=\sig$. This is a general description of stochastic quantum evolution, and will allow us to utilize the monotonicity of relative entropy to obtain a second law-like inequality. 

\subsection{Microcanonical ensemble model}
Consider first an open quantum system but without any net exchange of energy or particle number with the surroundings, in some quantum state \rh. A natural reference state is the microcanonical ensemble density operator $\sig_{\text{m}}$. For this it is not enough if $\rho$ has expectation values $E(\rho) = \text{Tr}\{ \rho H \}$ and $N(\rho) =  \text{Tr}\{ \rho N \}$ that agree with $E(\sigma_\text{m})$ and $N(\sig_\text{m})$, but $E$ and $N$ must be strictly fixed so that the variances vanish. We denote these two conditions by $E(\rho)\equiv E(\sigma_\text{m})$ and $N(\rho)\equiv N(\sigma_\text{m})$.

Then $\sig_{\text{m}}$ is a maximally mixed state corresponding to a uniform distribution, $\sig_{\text{m}} = \text{diag}(1/D,1/D, ...)$ where $D$ is the dimension of the Hilbert space of accessible states. The relative entropy of $\rh$ with respect to $\sig$ is then
\begin{align}
    S(\rh\|\sig_{\text{m}}) = -S(\rh) + S(\sig_{\text{m}}) = -S(\rh) + \ln D.
    \label{eq:Relative_Entropy_Micro}
\end{align}
Applying the CPTP map $\mathcal{N}$ and using the monotonicity property \eqref{monotonicity} directly gives (using $\mathcal{N}(\sigma)=\sigma$)
\begin{equation}
\begin{split}
    \Delta S(\rh \| \sig_{\text{m}}) & = S(\mathcal{N}(\rh) \| \sig_{\text{m}}) -  S(\rh \| \sig_{\text{m}}) \\
    & = - S\big(\mathcal{N}(\rh)\big) + S(\rh) = - \Delta S(\rho) \leq 0, \label{eq:MicroSecondlaw}
\end{split}
\end{equation}
equivalent to the second law of thermodynamics in the microcanonical ensemble. Note that the CPTP map $\mathcal{N}$ that keeps $\sig_\text{m}$ as a steady state defines a unital map. Thus equation \eqref{eq:MicroSecondlaw} is in fact equivalent to equation \eqref{eq:UnitalEntropy}.

\begin{figure}[t!]	
	\centering
    \includegraphics[width=0.45\textwidth]{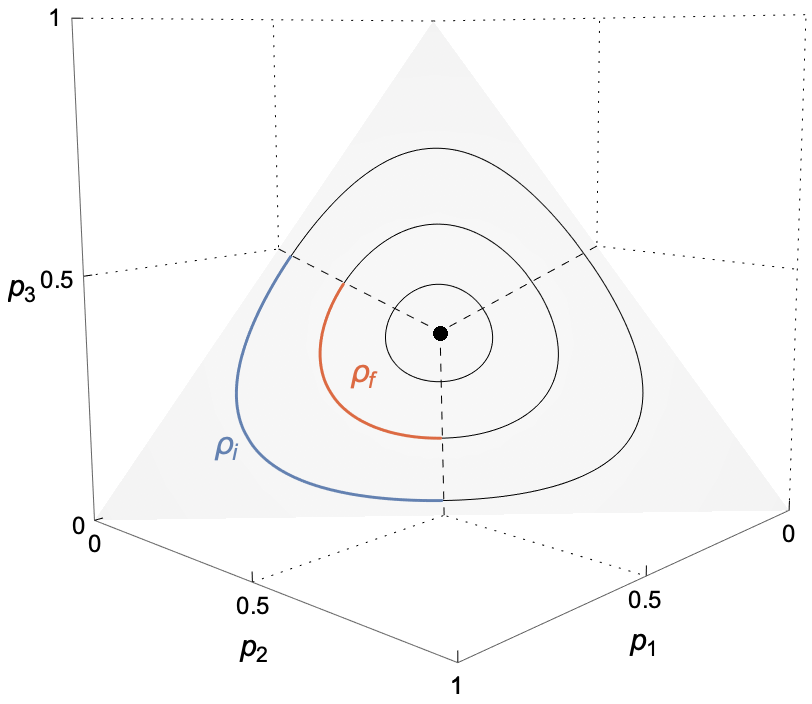}
    \caption{Curves of constant relative entropy $S(p\| \frac{1}{3}\mathds{1})$ or entropy $S(p)$ in the energy eigenbasis for a true distribution $(p_1,p_2,p_3)$ relative to a model uniform distribution $q_1=q_2=q_3=1/3$ (black dot). The dashed lines indicate permutation symmetry of the coordinates. Some initial state $\rh_i$ may evolve along the blue curve (reversible process), but can also evolve irreversibly to a final state $\rh_f$ on the red curve.} \label{fig:ConstRelEntropy}
\end{figure}

The formulation in terms of relative entropy can be illustrated geometrically, as seen in figure \ref{fig:ConstRelEntropy}. We consider a 3-state system and evaluate the states in the energy eigenbasis, where we assume no degeneracy of states for simplicity. The black contours indicate states of constant relative entropy with respect to a microcanonical model represented by the black dot in the middle. Clearly eq.\ \eqref{eq:Relative_Entropy_Micro} implies that these contours also represent constant entropy of the state $\rh$. The monotonicity now tells us that any initial state $\rh_i$, which is a point on, for example the blue contour, can only evolve along this contour or towards an inner contour (for example the red contour). It should be emphasized that the monotonicity may not necessarily tell us what exact state is taken by the system after the evolution, but rather on which contour the state may lie. Moreover, we can distinguish between reversible and irreversible processes, with a reversible process being characterized by a constant entropy during the evolution which translates to an evolution along one contour. In contrast, an irreversible process is then an evolution towards the center.

\subsection{Canonical ensemble model}
If the open quantum system \rh may exchange in addition to quantum information also energy with a heat bath, then a canonical thermal state is a suitable reference state, defined as
\begin{align}
    \sig_{\text{c}} = \frac{1}{Z}e^{-\beta H},
\end{align} 
where $Z = \text{Tr} \{e^{-\beta H} \}$ is the canonical partition function. The temperature $T=1/\beta$ is that of the heat bath and the condition $E(\rh) \equiv E(\sig_{\text{c}})$ is released, since energies are now allowed to fluctuate. Otherwise, $N(\rh) \equiv N(\sig_{\text{c}})$ still holds. Using the expression for the free energy $F(\sig_{\text{c}})=-(1/\beta) \ln Z  = E(\sig_{\text{c}}) - T S(\sig_{\text{c}})$ allows one to write the relative entropy as
\begin{equation}
    \begin{split}
    	S(\rh \| \sig_{\text{c}}) =& -S(\rh) + S(\sig_{\text{c}}) + \beta [E(\rh) - E(\sig_{\text{c}})].
    	\label{eq:thermal_rel_entropy_DelE}
    \end{split}
\end{equation}
After applying the CPTP map we find
\begin{equation}
    \begin{split}
        \Delta S(\rho \| \sigma_\text{c})  & = S(\mathcal{N}(\rh) \| \sig_{\text{c}}) - S(\rh \| \sig_{\text{c}}) \\ 
        &=  - S\big(\mathcal{N}(\rh)\big)  + S(\rh) + \beta [E(\mathcal{N}(\rh)) - E(\rh)] \\ 
        &= - \Delta S(\rho) + \beta \Delta E(\rho) \leq 0.
    \end{split} \label{eq:CanonicalSecondlaw}
\end{equation}
This means that the actual state $\rh$ may not diverge from the invariant equilibrium state $\sig_{\text{c}}$ under stochastic evolution in the sense of relative entropy, which is equivalent to the Clausius second law inequality. Stochastic evolution here is a quantum channel $\mathcal{N}$ which admits $\sig_{\text{c}}$ as a steady state. After the system has thermalized with the heat bath, its entropy and energy expectation value coincide with those of the canonical model. 

\subsection{Grand canonical ensemble model}

One may also consider an open quantum system, where in addition to quantum information also energy and particles may be exchanged with the environment or heat bath. The equilibrium state for this situation is described by the grand canonical ensemble $\sig_{\text{gc}}$
\begin{equation} \label{eq:GrandCan}
    \sig_{\text{gc}} = \frac{1}{Z} e^{-\beta (H - \mu N)},
\end{equation}
where $\mu=\alpha / \beta$ is the chemical potential and $Z$ is the grand canonical partition function. The condition $N(\rh) \equiv N(\sig_{\text{gc}})$ is now also released. Using an expression for the grand canonical potential, $\Omega = - (1/ \beta) \ln Z = E (\sig_{\text{gc}}) - T S (\sig_{\text{gc}}) - \mu N (\sig_{\text{gc}})$, one finds for relative entropy
\begin{equation}
\begin{split}
	S(\rh \| \sig_{\text{gc}}) =& -S(\rh) + S(\sig_{\text{gc}}) +\beta [E(\rh) - E(\sig_{\text{gc}})] \\
	&- \alpha[N(\rh) - N(\sig_{\text{gc}})].
	\label{eq:rel_entropy_DelE}
\end{split}
\end{equation}
If we calculate the difference in relative entropies after applying the CPTP map $\mathcal{N}$, we get an additional term in the Clausius relation due to particle exchange $\Delta N$
\begin{equation}
\begin{split}
    \Delta S(\rh \| \sig_{\text{gc}}) = & S(\mathcal{N}(\rh) \| \sig_{\text{gc}})  - S(\rh \| \sig_{\text{gc}}) \\
     = & - \Delta S(\rho) + \beta \Delta E(\rho)  - \alpha \Delta N(\rho) \leq 0.
\end{split} \label{eq:GrandCanonicalSecondlaw}
\end{equation}

In summary, the present section shows that the monotonicity of relative entropy for an open quantum system evolution implies a general form of the second law of thermodynamics, given an appropriate choice of an invariant thermal reference state. The advantage of this approach is that it is general; we can apply it to a wide range of thermodynamic situations and the density matrix \rh describes an arbitrary non-equilibrium state. Unlike standard thermodynamics, one does need to assume here quasi-stationary evolution from one equilibrium state to another.

Moreover, the presented arguments can be generalized to situations in which the model state $\sig$ is a non-equilibrium steady state (NESS). In these cases the condition $\mathcal{N}(\sig) = \sig$ still holds, such that a change in entropy is constrained from below. The main difference is that the remainder term is not of simple form and its meaning is not always clear. For further discussion see for example refs. \cite{Hatano2001,Breuer2010}.

\section{A local second law from relative entropy} 
\label{sec:LocalSecondLaw}

In this section we shall be concerned with generalizing the results of section \ref{sec:thermosecondlaw} to a relativistic quantum field theory. This can be done in several ways. The first and most direct application of the relations derived in section \ref{sec:thermosecondlaw} are for a global description of the field theory. This could be an infinitely large space, but also a finite spatial volume with appropriate boundary conditions (such as periodic boundary conditions). More interesting for applications to understand relativistic fluids are local descriptions, to which we turn afterwards. In particular, for a relativistic quantum field theory one can not only consider global time evolution, but one may also define evolution operators that propagate a state locally from one Cauchy surface to the next.

\subsection{Global time evolution}

Let us first consider the overall time evolution of a quantum field theory. To generalize the results of section \ref{sec:thermosecondlaw} we need to assume a coupling to some bath with exchange of quantum information and possibly also exchange of energy and particle number. A global evolution with time can be considered for all of space. Oftentimes, this is then an infinite volume, but one may also make the spatial volume finite by considering for example a generalized torus with periodic boundary conditions. The time evolution of the ``open quantum field theory'' is then given by a CPTP map $\mathcal{N}$, precisely as it has been discussed in the previous section. Accordingly one also obtains second law-type relations formulated with relative entropy as in eq.\ \eqref{eq:MicroSecondlaw} when only quantum information is being exchanged or in eq.\ \eqref{eq:CanonicalSecondlaw} with energy exchange or in eq.\ \eqref{eq:GrandCanonicalSecondlaw} with energy and particle number exchange, respectively.

\subsection{Local thermal equilibrium approximation} \label{sec:LocalThermalEquilibrium}

Before we generalize the results discussed in section \ref{sec:thermosecondlaw} to local relations, let us digress for a moment and consider more broadly the relation and interplay between local thermal equilibrium, fluid dynamics and locality in the context of a relativistic quantum field theory.

Relativistic fluid dynamics uses the concepts of thermal equilibrium not only in a global sense, i.\ e.\ for the entire set of quantum fields at some instance in time, but also {\it locally}, at a given point $x$ in space and time and a neighborhood around it. This brings new elements and features into the discussion.

Let us first emphasize that local thermal equilibrium is typically used as an approximate concept. It holds to lowest order in an expansion in gradients of fluid velocity, temperature etc., such that it becomes exact and equal to global thermal equilibrium when these gradients are absent. The lowest order of this derivative approximation leads to ideal fluid dynamics, with a corresponding form of the energy-momentum tensor and conserved particle current. On the other side, corrections to this ideal fluid limit are often sizeable and need to be taken into account. 

Dissipative, relativistic fluid dynamics exists in different forms. The dynamical variables may be the fields describing thermal equilibrium (fluid velocity, temperature, chemical potentials) \cite{Landau1987} or there might be additional fields that vanish in global equilibrium such as the shear stress $\pi^{\mu\nu}$, bulk viscous pressure $\pi_\text{bulk}$, and diffusion current $\nu^\mu$ in Israel-Stewart theory \cite{Israel1979}. Often the fields correspond to the degrees of freedom of the conserved energy-momentum tensor and conserved particle current, but in principle also other fields could appear.\footnote{A common element of all such fluid approximation is that they have much less degrees of freedom than a general out-of-equilibrium quantum state, highlighting again the approximation character of the description.}  

An interesting new feature of a local approximation is that it can neglect some non-local information of the quantum state such as entanglement between different spatial regions. It is conceivable that local observables in some region are well described by several quantum field theoretic states or density matrices, but that these differ in their global properties. As a simple example, degrees of freedom at two points or in two subsystems $A$ and $B$ are in general described by a density matrix $\rh_{AB}$, while local observables on either $A$ or $B$ are equally well described by the product of reduced density matrices $\rh_A \otimes \rh_B = \text{Tr}_B\{\rho\} \otimes \text{Tr}_A \{\rho\}$, which neglects the entanglement between $A$ and $B$. It might be possible to understand a local equilibrium approximation, or more generally a local fluid approximation, as an approximation of this kind: it works well for local observables but neglects non-local entanglement (and the associated correlations) to some extent.

Let us now formulate the above idea more concretely. We consider a quantum field theoretic state described by some density matrix \rh on a Cauchy hypersurface $\Sigma$ (for example the $d-1$ dimensional hypersurface of constant time t). We take this state $\rho$ to be out-of-global-equilibrium.\footnote{The following somewhat informal discussion will use the reduced density matrix for a spatial region in a quantum field theory. We note that this may not be well defined from a mathematical point of view. For example, as discussed in section \ref{sec:level1}, the corresponding von Neumann (entanglement) entropy is divergent. Ultimately we want to use relative entropies which can alternatively, and rigorously, be defined in terms of modular theory \cite{Araki1977}.}

Now let us concentrate on some subregion $A$ of $\Sigma$, say a ball of radius $R$ around some point $\vec x_A$ at time $x_A^0$. Local observables in this region can be described by the reduced density matrix
\begin{equation}
    \rh_A = \tr_{\bar{A}} \{\rh \},
\end{equation}
where the partial trace goes over the complement region $\bar{A}$ such that $\Sigma = A \cup \bar{A}$ and $\bar{A} \cap A = \emptyset$.
We may now consider a {\it global} equilibrium state \sig, specified by $\beta^\mu$ and $\alpha$, as defined in eq. \eqref{eq:DefThermodynamicFields}. We may also similarly consider the reduced density matrix of this state to the region $A$, 
\begin{equation}
    \sig_A = \tr_{\bar{A}} \{\sig \}.
\end{equation}
One may now say that the (non-equilibrium) state \rh is in {\it local} thermal equilibrium in the region $A$ around $\vec x_A$ (with local $\beta^\mu$ and $\alpha$) when the two reduced density matrices agree, $\rh_A = \sig_A$. Note that these two statements may depend somewhat on the size of the region $A$ around $x_A$, i.\ e.\ the radius $R$. In practice, this size is taken to be small enough from a macroscopic point of view such that its precise value is not relevant, while from a microscopic point of view it has to be large enough, for example compared with possible UV regulator scales of the (effective) quantum field theory. The concept of such an intermediate scale, which defines a {\it fluid cell}, also appears in other formulations of fluid dynamics, for example in the context of kinetic theory \cite{Landau1987}. 

One may actually quantify how well a local thermal equilibrium description (as introduced above) works in terms of the relative entanglement entropy $S(\rh_A \| \sig_A)$. In particular, \sig is locally a good model for the state \rh in the region $A$ when they become \textit{locally indistinguishable}, i.e.
\begin{equation} \label{eq:RelEntropyToZero}
    S(\rh_A \| \sig_A) \to 0.
\end{equation}
This is a purely information theoretic criterion for a state to be {\it locally} of thermal equilibrium form (in the region $A$). Of course, when \rh is globally out-of-equilibrium, one has necessarily $S(\rh_A \| \sig_A) > 0$ once the region $A$ is large enough. This shows again the necessity for an intermediate scale (a fluid cell size) where a local equilibrium description can work.

It may also be possible that a local equilibrium description works everywhere on the hypersurface $\Sigma$ in the sense that one can assign to each point $x$ local values $\beta^\mu(x)$ and $\alpha(x)$ in the sense described above and such that \eqref{eq:RelEntropyToZero} is fulfilled for a convenient neighborhood of the point $x$. This does not imply that \rh is itself a (global) equilibrium state, and in fact it cannot be if $\beta^\mu(x)$ is not a Killing field and $\alpha(x)$ is not constant, respectively. This situation corresponds to an ideal fluid approximation being approximately valid. In this case one can also find a globally defined state \sig given by 
\begin{equation}\label{eq:ledo}
    \sig  = \frac{1}{Z} \exp \Big[ -\int_{\Sigma (\tau)} d \Sigma_\mu \big\{ \beta_\nu(x) T^{\mu \nu}(x)   + \alpha(x) N^\mu(x) \big\} \Big],
\end{equation}
where 
\begin{align}
    Z= \text{Tr}\Big\{\exp \Big[-\int_{\Sigma (\tau)} d \Sigma_\mu \{ \beta_\nu T^{\mu \nu} + \alpha N^\mu \}\Big]\Big\}
\end{align} 
is a generalized partition function, such that for any fluid cell $A$, eq.\ \eqref{eq:RelEntropyToZero} is fulfilled. This does not imply that $\rh = \sig$ on a global level, however. It is even conceivable that \rh is a pure state, while \sig is obviously mixed. The states differ in their global properties while agreeing locally. 

So far we have concentrated on situations where the local description used only the thermodynamic parameters $\beta^\mu$ and $\alpha$. Beyond this it may sometimes be necessary to use a more complex local approximation, for example to represent locally the entire energy momentum tensor, beyond its ideal fluid components, faithfully. In a spirit similar to the above discussion one may say that a local fluid approximation state \sig is a good description when the corresponding reduced density matrices $\rh_A$ and $\sig_A$ agree such that eq. \eqref{eq:RelEntropyToZero} is fulfilled. We will discuss a class of such states \sig, for which local equilibrium states as in eq.\ \eqref{eq:ledo} and global equilibrium states are a subclass, in section \ref{sec:BathGeneraExponential}.

An interesting and important question is how the local approximate states evolve in time. For the true state \rh, and an isolated situation, the time evolution is unitary. In contrast, a class of states \sig that approximates \rh locally but differs from it globally, does not have to evolve in a unitary way. It is conceivable that after unitary time evolution of some local equilibrium state as in eq. \eqref{eq:ledo} it is not part of these class of states any more (i.\ e.\ it cannot itself be written as in equation \eqref{eq:ledo}). At the same time it may be possible to represent the full state \rh also after some time evolution, again {\it locally} by states of the form \eqref{eq:ledo}. It is intuitively clear that the states \sig used for a local approximation have a sort of coarse-grained evolution. For this time evolution, the quantum information does not have to be conserved, because non-local entanglement is at least partly dropped. As the trace is preserved and the density matrix must remain positive, the coarse-grained evolution should be a (C)PTP map.

In the present work we do not attempt to develop such a coarse-grained description of dynamical evolution in more detail. Instead we consider a different but closely related situation where quantum information can also get lost, but now through the coupling to an external local ``bath fluid''. Formally we deal then with an open quantum system for which the time evolution is again not unitary. When the bath fluid is not described explicitly, but is instead effectively ``integrated out'', the quantum fields we consider evolve themselves by CPTP maps. In the following we will develop these descriptions in more detail, with different scenarios for the bath fluid and its coupling to the quantum fields under consideration. 

\subsection{Local evolution and double light cone}

Besides global time evolution, one may in a relativistic quantum field theory also consider more general evolution operators that evolve the state from one Cauchy hypersurface to the next. These Cauchy surfaces must have normal vectors that point into a time-like (or, as a limit, light-like) direction and they should be ordered such that the evolution does nowhere go backwards in time. Otherwise, they can be chosen quite freely.

For our purpose this is intriguing, because we are interested in a local form of the second law. The strategy is therefore to consider a series of hypersurfaces that differ only in a well localized region in space, so that the evolution is essentially local. 

In the following we will investigate how eq.\ \eqref{eq:LocalSecondLaw} can be understood from a quantum field and quantum information theoretic point of view. After integration, eq.\ \eqref{eq:LocalSecondLaw} states that in a certain region of space-time $\Omega$ entropy can only increase, but not decrease,
\begin{equation}
\int_\Omega d^dx \sqrt{g}\,  \nabla_\mu s^\mu(x) = \oint_{\partial \Omega} d \Sigma_\mu s^\mu(x) \geq 0.
\label{eq:LocalSecondLawIntegralFormulation}
\end{equation}
We use here the (hyper-)surface element
\begin{equation}
    d\Sigma_\mu = d^{d-1}y\ \sqrt{h}\ n_\mu,
\label{eq:surfaceElementwithNormalVector}
\end{equation} where $n_\mu$ is a local unit vector normal to the surface and $h=|\det h_{\mu\nu}|$ is the determinant of the induced metric on the hypersurface. Alternatively, in terms of differential forms one may write,
\begin{equation}
    d\Sigma_\mu = \frac{1}{(d-1)!}\sqrt{g} \epsilon_{\mu\nu_1\cdots \nu_{d-1}} dx^{\nu_1}\wedge \cdots \wedge dx^{\nu_{d-1}},
    \label{eq:surfaceElementDF}
\end{equation}
where $g=-\det g_{\mu\nu}$ is the determinant of the metric. 

Let us remark here on some subtleties in the orientation of the normal vector in eq.\ \eqref{eq:surfaceElementwithNormalVector} in a space with metric signature $(-,+,+,+)$. For a part of the closed surface where the normal vector is space-like, the orientation is unambiguously taken as pointing to the outside. For parts where the normal vector is time-like, it must then be taken such that the normal vector is orientated inwards (see refs. \cite{Wald1984} \S B.2, \cite{Gourgoulhon2013} \S 16 and also \cite{Lee2012} for mathematical details). For a closed surface that consists of two Cauchy surfaces, this means that $n^0<0$ on the future lying Cauchy surface and $n^0>0$ for the past. This will have to be taken into account below. Specifically, when we integrate over a Cauchy hypersurface as in eq.\ \eqref{eq:ledo}, we usually assume $n^\mu$ to be future oriented such that $n^0 >0$, similar to a fluid velocity.

\begin{figure}[t!]	
        \centering
        \includegraphics[width=8.6cm]{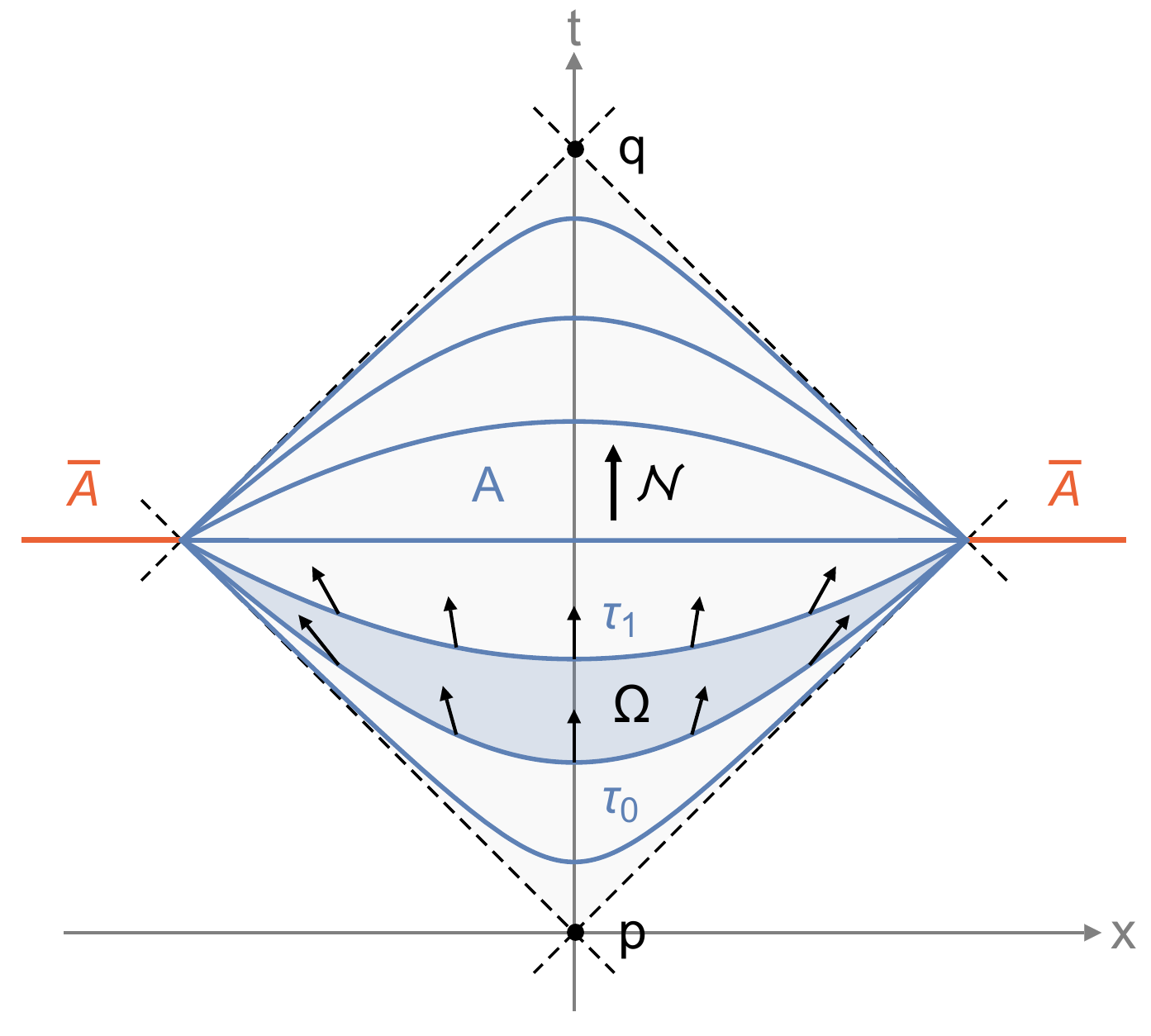}
        \caption{Double light cone region bounded by one light cone starting at the point $p$ and one ending at the point $q$. We consider a series of Cauchy surfaces that evolve within the light cone region but remain fixed in the region where the two cones intersect, as well as in the outside region. Different such Cauchy surfaces are labeled by the parameter $\tau$. Then the restriction of a surface $\Sigma (\tau)$ to the double line cone is denoted $A (\tau)$, and the complement region $\bar{A}(\tau)$. The operator $\mathcal{N}$ evolving the density matrix from one Cauchy surface to the next is unitary for an isolated system but a more general CPTP map for quantum fields coupled locally to some external bath fluid. Two consecutive hypersurfaces enclose a space-time volume $\Omega$, where the small black arrows indicate the normal vectors $n^{\mu}(x)$ of $\Sigma(\tau)$.} \label{fig:DoubleLightCone}
\end{figure}

As a direct consequence of the divergence theorem, eq.\ \eqref{eq:LocalSecondLaw} implies \eqref{eq:LocalSecondLawIntegralFormulation} for any region $\Omega$ with boundary $\partial \Omega$.
On the other side, eq.\ \eqref{eq:LocalSecondLawIntegralFormulation} also implies eq.\ \eqref{eq:LocalSecondLaw} if we can prove it for some space-like region around the point $x$ that can be made arbitrarily small. In the following we shall choose a particular geometry for such a space-time region, namely the double light cone as illustrated in Fig.\ \ref{fig:DoubleLightCone}.  This geometry, that is bounded by two light cones, one originating at a point $p$ in the past of $x$ and one ending at a point $q$ in its future, has the advantage that the spatial boundary is just the two-dimensional intersection of the cones. The past light cone originating from $p$ forms a $(d-1)$-dimensional light-like part of the boundary that can also be understood as an initial hyper surface, while the future light cone ending at $q$ can also be understood as a final hyper surface for the evolution inside the double light cone region itself. As we will see, this has great advantages for the quantum field theoretic discussion, see also, for example, ref.\ \cite{Haag1996}. 
 
For a situation where the quantum fields are in isolation, i.\ e.\ without any interaction with an external bath fluid,  one would have unitary time evolution in the entire system but also locally within the double light cone. Instead, if the system is not isolated one can still define density matrices for the different hyper surfaces and evolution operators between them, even though they are not unitary any more. This is the situation we want to address here.

More formally, the coupling to the bath fluid is supposed to be via a convenient local interaction term, even though we do not specify the latter explicitly.

Within the double light-cone region as well as outside of it, we choose a one parameter family of $(d-1)$-dimensional spacelike hypersurfaces $\Sigma(\tau)$ with timelike unit normal $n_\mu(x)$, defined as a foliation of space-time where $\tau$ can be thought of as a generalized time coordinate. The manifold where the two light cones intersect correspond to a set of {\it fixed points}, i.\ e.\ it is part of all $\Sigma(\tau)$. For $d=1+1$ dimensions (as shown in Fig.\ \ref{fig:DoubleLightCone}) the intersection of the two light cones has just two points while it is a 2-sphere for $d=1+3$ dimensions. Outside of the double light cone region the Cauchy surfaces do not change with $\tau$ so that all evolution happens actually within this region. The restriction of the spacelike hypersurfaces $\Sigma(\tau)$ to the double light cone region itself will be called $A(\tau) \subseteq \Sigma(\tau)$. Similarly, we denote the corresponding complement region on $\Sigma(\tau)$ by $\bar A$. 

We define the actual states of our system on these surfaces by the family of density operators $\rh(\tau)$, and also define reference states as $\sig(\tau)$ which we will later specify to be some form of equilibrium state in analogy to section \ref{sec:thermosecondlaw}. One should note that the actual state of the system $\rho$ is arbitrary in the following and may be of non-equilibrium form. For hypersurfaces that are chosen such that the intersection of the two light cones corresponds to fixed points, the dynamics inside are isolated from the outside in the sense that there is no transfer of quantum information through the boundaries during the evolution. The only quantum information from outside being able to affect physics within the double light cone is encoded on the past boundary as initial conditions. This ensures that any entropy production will be solely within this region and will not be affected by the exterior.

As mentioned before, we will consider here an open quantum system evolution, where the sector of the theory we consider is coupled to some ``bath fluid''. This bath fluid can either be in a global equilibrium state, a situation we will discuss next in subsection \ref{sec:BathGlobalEquilibrium}, or it could be in a more general local equilibrium state, which we constructed in subsection \ref{sec:LocalThermalEquilibrium} and which will be discussed in subsection \ref{sec:BathLocalEquilibrium}. In subsection \ref{sec:BathGeneraExponential} we will then consider an even more general situation. In any case, the evolution of the quantum field theoretic state in contact with the bath fluid from a density operator $\rh(\tau_0)$ to some other state $\rh(\tau_1)$ is described by a completely positive trace-preserving (CPTP) map,
\begin{equation}
    \mathcal{N}: \quad \rho(\tau_0) \rightarrow \mathcal{N} ( \rh(\tau_0) )=\rho(\tau_1),
\end{equation}
where the particular map $\mathcal{N}$ depends on the initial and final hypersurfaces. The same map is being applied to the reference state \sig, however the latter will typically be chosen such that it is invariant or covariant under $\mathcal{N}$.

\subsection{Coupling to a ``bath fluid'' in global equilibrium}\label{sec:BathGlobalEquilibrium}
We now specify the ``bath fluid'' to be a fluid in global thermal equilibrium governed by $\beta^\nu=u^\nu/T$, the ratio of fluid velocity and temperature, and similarly $\alpha=\mu/T$, the ratio of chemical potential and temperature. We may quantify the coupling between the fluid we are actually interested in and the bath fluid by some interaction parameter $\lambda$. An example could be a field theory of electrons and positrons coupled to a bath of electromagnetic radiation through the usual coupling strength $e$. (In that case the chemical potential would vanish.) Another example would be the field for up quarks coupled to a bath of down quarks, strange quarks, gluons etc. We expect dissipative effects to be strong for large $\lambda$, whereas $\lambda \to 0$ leads to unitary time evolution for the sector of the theory we describe explicitly.\footnote{We note that in ref.\ \cite{Yao2019} a similar open quantum system setup as well as monotonicity of relative entropy are  considered in the context of quarkonium formed in heavy ion collisions and assumed to be weakly coupled to a locally equilibrated hot quark-gluon plasma.}

In the following we will not describe the bath fluid explicitly but keep its presence in mind. For the sector of the theory we describe explicitly, we compare two states or density matrices, $\rho(\tau)$ and $\sigma(\tau)$. While $\rho(\tau)$ is generically out-of-equilibrium, we shall assume in the following that $\sigma(\tau)$ is a reference state in global thermal equilibrium with the bath fluid. We can then directly specify the density operator of the latter on some hypersurface $\Sigma(\tau)$,
\begin{equation}\label{eq:gedo}
    \sig  = \frac{1}{Z} \exp \Big[ -\int_{\Sigma (\tau)} d \Sigma_\mu \big\{ \beta_\nu T^{\mu \nu}   + \alpha N^\mu \big\} \Big],
\end{equation}
where 
\begin{align}
    Z= \text{Tr}\Big\{\exp \Big[-\int_{\Sigma (\tau)} d \Sigma_\mu \{ \beta_\nu T^{\mu \nu} + \alpha N^\mu \} \Big]\Big\}
\end{align} 
is the thermal partition function. This definition together with the conditions \eqref{eq:KillingEquation} corresponds to the covariant generalization of a (time independent) equilibrium state. It also provides a unique fluid frame with time-like fluid velocity $u^\mu$ in the direction of the Killing field $\beta^\nu$, such that in an equilibrium without rotation or acceleration one may write
\begin{equation}
\begin{split}
    &s^\mu = s u^\mu, \ \ \  \ \ \ T^{\mu \nu} = \varepsilon u^\mu u^\nu + p \Delta^{\mu \nu},  \\ &\text{and}  \ \ \ \ \ \  N^\mu = n u^\mu, \label{eq:EquilibriumQuantities} 
\end{split}
\end{equation}
where $s$ is  defined as the entropy density, $\varepsilon$ is the energy density and $n$ is the particle density of the fluid. Note that $T^{\mu \nu}$ and $N^\mu$ are expectation values here, as opposed to being operators in eq.\ \eqref{eq:gedo}. 

We now wish to formulate a second law in local form. A problem to overcome here is that the total von Neumann entropy $S(\rho)$ of an arbitrary state cannot easily be written as an integral over some local entropy current. This is because outside of equilibrium entropy is not necessarily extensive, i.\ e.\ for a fluid proportional to the volume. However, the entropy of the equilibrium state \sig may be expressed in such a way and one can write using eq.\ \eqref{eq:gedo} 
\begin{equation}
    \begin{split}
        S(\sig) &= - \text{Tr}\{ \sig \ln \sig \} \\
        &= \ln(Z)+ \int d \Sigma_\mu \{ \beta_\nu T^{\mu \nu}(\sig) +\alpha N^\mu(\sig) \}  \\
        &= -\int d\Sigma_\mu s^{\mu}(\sig).
    \end{split} \label{eq:Ssigma}
\end{equation}
(The minus sign in the last line arises because we take $d\Sigma^\mu$ and $s^\mu$ to be future-oriented with positive time components $d\Sigma^0$ and $s^0$, and work with metric signature $(-,+,+,+)$.) We have used here the expectation values
\begin{equation}
T^{\mu\nu}(\sigma) = \text{Tr}\{ \sigma T^{\mu\nu} \}, \quad\quad\quad N^{\mu}(\sigma) = \text{Tr}\{ \sigma N^{\mu} \},
\end{equation}
which we take to be renormalized, such that they vanish in vacuum where $T=\mu=0$. We are also using here that the logarithm of the partition function (the Schwinger functional for vanishing source) can be written as
\begin{equation}
W = \ln(Z) = - \int d\Sigma_\mu  \{ p \beta^\mu \},
\end{equation}
where $p$ is the pressure and we are again assuming a renormalization such that $p=0$, and accordingly $Z=1$, in vacuum where $T=\mu=0$. The thermal entropy current can be written as
\begin{equation}
s^\mu = - \beta_\nu T^{\mu\nu} - \alpha N^\mu + p \beta^\mu,
\label{eq:smuequilibrium}
\end{equation}
and using the relation $\epsilon+p=sT + \mu n$ one can see that for an ideal fluid where $T^{\mu\nu}=(\epsilon+p)u^\mu u^\nu + p g^{\mu\nu}$ and $N^\mu = n u^\mu$, eq.\ \eqref{eq:smuequilibrium} agrees indeed with the usual definition $s^\mu=s u^\mu$. 

The relative entropy between an arbitrary state \rh and the global equilibrium state \sig at some time parameter $\tau$ reads
\begin{equation} 
    \begin{split}
        S(\rh \|\sig) & =  \text{Tr} \left\{ \rh\big( \ln(\rh) - \ln(\sig)\big) \right\}  \\
         = &-S(\rh) + \ln(Z) + \text{Tr}\Big\{ \rh \int d\Sigma_\mu \big( \beta_\nu T^{\mu \nu }  +\alpha N^\mu \big)\Big\} \\
        = & -S(\rho) +\int d\Sigma_\mu \Big\{ -s^\mu (\sig)  \\ 
        &+ \beta_\nu \big[T^{\mu \nu}(\rh) -T^{\mu \nu}(\sig)\big] + \alpha \big[N^\mu(\rh) -N^\mu(\sig)\big] \Big\}.
        \end{split}
        \label{eq:relentropyrhosigmalocallythermal}
\end{equation} 
Note that the right hand side of \eqref{eq:relentropyrhosigmalocallythermal} contains the part $-S(\rho)$ that is defined in a non-local way, and a local part written as an integral over the Cauchy hypersurface. The integral is here over all of $\Sigma=A \cup \bar{A}$. However, the time evolution we consider is such that it takes place only in $A$, while the part of the Cauchy surface denoted $\bar A$ remains stationary. 

In a next step we may consider the difference of relative entropies between two Cauchy surfaces,
\begin{equation}
\Delta S(\rh \|\sig) = S(\rh(\tau_1) \|\sig(\tau_1)) - S(\rh(\tau_0) \|\sig(\tau_0)).
\label{eq:DeltaSRelDef}
\end{equation}
We assume here that the time evolution is such that the global thermal equilibrium state $\sigma$ is stationary, i.\ e.\ it remains to be of the form \eqref{eq:gedo}, even though $\sig(\tau_1)$ and $\sig(\tau_0)$ are defined on different Cauchy surfaces. In contrast, the state $\rho$ is not stationary, so that $\rho(\tau_1) = \mathcal{N}(\rho(\tau_0))$ and $\rho(\tau_0)$ are different states. As a consequence of the coupling to the external bath fluid, the evolution operator $\mathcal{N}$ is in general not unitary but a CPTP map. 

From the monotonicity property of relative entropy under CPTP maps it follows that 
\begin{equation}
\Delta S(\rho \| \sigma) \leq 0, 
\label{eq:monotonicityRE}
\end{equation}
where the equality is for vanishing coupling to the bath fluid, $\lambda=0$, corresponding to unitary time evolution.

Using \eqref{eq:relentropyrhosigmalocallythermal} we can rewrite the difference of relative entropies as
\begin{equation}
\begin{split}
&  \Delta S(\rho \| \sigma) = - \Delta S(\rho) - \oint\limits_{A(\tau_1) \cup A(\tau_0)} d\Sigma_\mu \Big\{ - s^\mu(\sigma) \\
& \quad+ \beta_\nu \big[T^{\mu \nu}(\rh) -T^{\mu \nu}(\sig)\big] + \alpha \big[N^\mu(\rh) -N^\mu(\sig)\big] \Big\}.
\end{split}
\label{eq:DeltaSRelThermal}
\end{equation}
The integral in \eqref{eq:DeltaSRelThermal} is now along a closed surface, and for such surface integrals we adopt the convention explained below eq.\ \eqref{eq:LocalSecondLawIntegralFormulation} (this explains the additional minus sign in front of the integral). We note in particular that in the difference, $\Delta S(\rho \| \sigma)$, contributions to the integral in \eqref{eq:relentropyrhosigmalocallythermal} from the region $\bar A$ outside of the double light cone have dropped out. 

We also use in \eqref{eq:DeltaSRelThermal} the difference of entropies $\Delta S(\rho) = S(\rho(\tau_1)) - S(\rho(\tau_0))$. While the entropy $S(\rho)$ is not generically local, any change in entropy is due to interactions with the bath fluid. Assuming that these processes are local in space and time allows to write
\begin{equation}
\Delta S(\rho) = S(\rho(\tau_1)) - S(\rho(\tau_0)) = \int_\Omega d^d x \sqrt{g} \; \mathfrak{s}(\rho)(x),
\label{eq:DeltaSLocal}
\end{equation}
where the integral goes over the space-time region $\Omega$ between the two Cauchy surfaces. The local form in \eqref{eq:DeltaSLocal} is also further supported by the fact that the Cauchy surfaces $\Sigma$ can evolve quite arbitrarily and may only change within some region.

The remaining terms in \eqref{eq:DeltaSRelThermal} can be rewritten by using the divergence theorem such that we obtain
\begin{equation}
\begin{split}
\Delta S(\rho \| \sigma) = \int\limits_\Omega d^d x \sqrt{g} \Big\{ & - \mathfrak{s}(\rho) - \beta_\nu \nabla_\mu T^{\mu\nu}(\rho) \\
& - \alpha \nabla_\mu N^\mu(\rho) \Big\} \leq 0.
\end{split}
\label{eq:DeltaSRelThermal2}
\end{equation}
We have used here that $\sigma$ is a global thermal equilibrium state such that its entropy current is conserved, $\nabla_\mu s^\mu(\sigma)=0$, and similarly also its energy momentum tensor and particle number current. Moreover, $\beta^\mu$ and $\alpha$ obey \eqref{eq:KillingEquation}. 

Because \eqref{eq:DeltaSRelThermal2} must be obeyed for any choice of the Cauchy surfaces $\Sigma$, we can conclude that the local relation
\begin{equation}
\mathfrak{s}(\rho) + \beta_\nu \nabla_\mu T^{\mu\nu}(\rho) + \alpha \nabla_\mu N^\mu(\rho) \geq 0,
\end{equation}
must hold, as well. This can be seen as a local version of the second law of thermodynamics in the present situation. Specifically, it is the local and differential version of eq.\ \eqref{eq:GrandCanonicalSecondlaw}. In particular, we find that a local version of the second law can be formulated in terms of relative entropy.

Let us emphasize again that the change in relative entropy in the present context is due to interactions with the bath fluid. Similar to eq.\ \eqref{eq:DeltaSLocal}, one can for local interactions also write the change in relative entropy between two Cauchy surfaces in a local way,
\begin{equation}
    \Delta S(\rho \| \sigma) = \int_\Omega d^dx \sqrt{g} \; \mathfrak{s}(\rho \| \sigma)(x) \leq 0.
    \label{eq:DeltaSrhosigmaLocal}
\end{equation}
Because this should hold for arbitrary Cauchy surfaces we find for the local ``production of relative entropy''
\begin{equation}
\mathfrak{s}(\rho \| \sigma)(x) \leq 0.
\end{equation}
Relative entropy can in some space-time volume only decrease, so that the states become less distinguishable, and not increase.

While we have now found a local formulation of the second law based on relative entropy, it would actually be interesting to go one step further and formulate this with {\it relative entanglement entropy} instead. The analog of \eqref{eq:DeltaSRelDef} is then 
\begin{equation}
    \Delta S(\rho_A \| \sigma_A) = S(\rh_A(\tau_1) \|\sig_A(\tau_1)) - S(\rh_A(\tau_0) \|\sig_A(\tau_0)).
    \label{eq:DeltaSRelEntanglementDef}
\end{equation}
We use here the reduced density matrices for the double light cone region
\begin{equation}
    \rho_A = \text{Tr}_{\bar A} \{ \rho \}, \quad\quad\quad \sigma_A = \text{Tr}_{\bar A} \{ \sigma \}.
\end{equation}
The partial traces are over the complement region $\bar A$ outside of the double light cone where the Cauchy surfaces remain stationary.

The reduced density matrices evolve according to modified evolution operators $\mathcal{N}_A$ such that $\rho_A(\tau_1) = \mathcal{N}_A(\rho_A(\tau_0)) = \mathcal{N}_A(\text{Tr}_{\bar A}\{\rho(\tau_0) \})  = \text{Tr}_{\bar A}\{\mathcal{N}(\rho(\tau_0))\} = \text{Tr}_{\bar A}\{ \rho(\tau_1) \}$. Because the double light cone region has fixed spatial boundaries, a non-vanishing difference in eq.\ \eqref{eq:DeltaSRelEntanglementDef} can only be a result of interactions with the bath fluid. It follows from monotonicity of relative entropy that $\Delta S(\rho_A \| \sigma_A) \leq 0$, with equality for vanishing interaction $\lambda=0$. If these interactions with the bath fluid are again local, it should be possible to write in analogy to \eqref{eq:DeltaSrhosigmaLocal}
\begin{equation}
    \Delta S(\rho_A \| \sigma_A) = \int_\Omega d^dx \sqrt{g} \; \mathfrak{s}(\rho_A \| \sigma_A) \leq 0.
\end{equation}
Moreover, even though we will not formally prove this, it is highly plausible that the local changes in relative entropy and relative entanglement entropies agree, 
\begin{equation}
    \mathfrak{s}(\rho_A \| \sigma_A) = \mathfrak{s}(\rho \| \sigma), 
\end{equation}
and as a consequence also 
\begin{equation}
    \Delta S(\rho_A \| \sigma_A) = \Delta S(\rho \| \sigma).
\end{equation}
This is quite an interesting possibility, because it allows to formulate the local version of the second law of thermodynamics not only in terms of relative entropy, but also in terms of relative entanglement entropy.

\subsection{Coupling to a ``bath fluid'' in local equilibrium}\label{sec:BathLocalEquilibrium}

We now aim to generalize somewhat the physics setting and allow for the bath fluid to deviate from global thermal equilibrium but assume it to be in {\it local} thermal equilibrium, instead. First of all, from the discussion in the previous subsection one expects that for the evolution within the double light cone region, only the local state therein is actually relevant. If the reduced density matrix $\sigma_A$ is actually the same as for a globally thermal state, the analysis of section \ref{sec:BathGlobalEquilibrium} goes through without essential modifications. 

Here we generalize this discussion to a situation where also within the double light cone region itself the bath fluid is not in equilibrium, but in a more general state. More specifically, we assume now that the bath fluid is such that an evolution map $\mathcal{N}$ is induced for the quantum fields we study that leaves a {\it local} equilibrium state $\sigma$ invariant. The latter is written as in eq.\ \eqref{eq:gedo}, but now $\beta^\mu(x)$ is not assumed to be a Killing vector field and $\alpha(x)$ is not taken to be constant. Instead we assume that $\beta^\mu(x)$ and $\alpha(x)$ are just some given fields, or functions of space and time. We assume that the map $\mathcal{N}$ that propagates the state from one hypersurface to the next is such that $\sigma$ remains to be of the particular form  in eq.\ \eqref{eq:gedo}.\footnote{As a side remark we note that a similar class of states appears also in the Zubarev approach where a non-equilibrium state is constructed by maximizing entropy on a given (and fixed) Cauchy surface given certain constraints involving expectation values of energy and momentum \cite{Zubarev1979, Becattini2019}.}

Interestingly, for a so-defined {\it local} equilibrium state, the first two lines of eq.\ \eqref{eq:Ssigma} remain valid, i.\ e.\ one can write the von Neumann entropy of such a state in terms of the partition function and the expectation values of energy-momentum tensor and particle current. Also, if the logarithm of the partition function, the Schwinger functional, is local or extensive,
\begin{equation}
W = \ln(Z) = - \int d\Sigma_\mu w^\mu(\sigma),
\label{eq:PartitionFunctionLocalEq}
\end{equation} 
one can introduce through eq.\ \eqref{eq:Ssigma} and the identification
\begin{equation}
s^\mu(\sigma) = - \beta_\nu T^{\mu\nu}(\sigma) - \alpha N^\mu(\sigma) + w^\mu(\sigma),
\label{eq:entropyCurrentLocalEq}
\end{equation}
an entropy current, as in the third line of eq.\ \eqref{eq:Ssigma}. Arguments for the applicability of eq.\ \eqref{eq:PartitionFunctionLocalEq} and the resulting entropy current \eqref{eq:entropyCurrentLocalEq} were recently given in ref.\ \cite{Becattini:2019poj}. Because the class of states introduced by this prescription is not in equilibrium when \eqref{eq:KillingEquation} is not fulfilled, eq.\ \eqref{eq:entropyCurrentLocalEq} is to be understood here as the definition of a non-equilibrium entropy current. The relation $\Delta S(\sigma) \geq 0$ can be written locally as $\nabla_\mu s^\mu(\sigma) \geq 0$.

As a check, for $w^\mu=p\beta^\mu$, eq.\ \eqref{eq:entropyCurrentLocalEq} gives indeed the right entropy current within first order relativistic fluid dynamics in the Landau frame \cite{Landau1987}. More generally, it would be good to check from the quantum field theory side whether the class of local equilibrium states in  eq.\ \eqref{eq:gedo} is a good approximation for certain out-of-global equilibrium situations.

In a next step one can consider the relative entropy of some generic state $\rho$ relative to the so-defined {\it local} equilibrium state $\sigma$. It is then not difficult to see that this relative entropy $S(\rho \| \sigma)$ can still be written as in eq.\ \eqref{eq:relentropyrhosigmalocallythermal}, of course with the difference that $\beta_\nu$ is not Killing and $\alpha$ is not constant any more.

In a subsequent step one may consider a difference of relative entropies on two Cauchy hypersurfaces as in eq.\ \eqref{eq:DeltaSRelDef}. By monotonicity of relative entropy this difference is non-negative, as expressed in eq.\ \eqref{eq:monotonicityRE}. Via eq.\ \eqref{eq:DeltaSRelThermal} and eq.\ \eqref{eq:DeltaSLocal} one is again lead to a relation that generalizes eq.\ \eqref{eq:DeltaSRelThermal2}, namely
\begin{equation}
    \begin{split}
        & \Delta S(\rho \| \sigma) = \int\limits_\Omega d^d x \sqrt{g} \Big\{ - \mathfrak{s}(\rho) + \nabla_\mu s^{\mu}(\sigma) \\
        & - \beta_\nu \nabla _\mu  \left[ T^{\mu \nu}(\rh) -T^{\mu \nu}(\sig) \right] - \alpha \nabla _\mu \left[ N^\mu (\rh) - N^\mu (\sig) \right] \\
        & -  (\nabla _\mu \beta_\nu)  \left[ T^{\mu \nu}(\rh) -T^{\mu \nu}(\sig) \right] - (\partial_\mu \alpha) \left[ N^\mu (\rh) - N^\mu (\sig) \right] \Big\}\\
        & \leq 0.
    \end{split}
\label{eq:localSecondLawLocalEquilibrium}
\end{equation}
One may use the definition of the entropy current \eqref{eq:entropyCurrentLocalEq} which allows us to simplify eq.\ \eqref{eq:localSecondLawLocalEquilibrium} to 
\begin{equation}
    \begin{split}
        & \Delta S(\rho \| \sigma) = \int\limits_\Omega d^d x \sqrt{g} \Big\{ - \mathfrak{s}(\rho) + \nabla_\mu w^{\mu}(\sigma) \\
         & - \beta_\nu \nabla _\mu   T^{\mu \nu}(\rh) - \alpha \nabla _\mu N^\mu (\rh)  \\
            & -  (\nabla _\mu \beta_\nu)  T^{\mu \nu}(\rh) - (\partial_\mu \alpha)  N^\mu (\rh)  \Big\} \leq 0.
        \end{split}
    \label{eq:localSecondLawLocalEquilibrium2}
\end{equation}
In a situation where the bath fluid exchanges no energy, momentum or particle number with the fields of interest, the second lines in \eqref{eq:localSecondLawLocalEquilibrium} as well as \eqref{eq:localSecondLawLocalEquilibrium2} drop out and we are left with
\begin{equation}
    \begin{split}
    \mathfrak{s}(\rho \| \sigma) = & - \mathfrak{s}(\rho) + \nabla_\mu s^{\mu}(\sigma) \\
        & -  \frac{1}{2}(\nabla_\mu \beta_\nu + \nabla_\nu\beta_\mu) \left[ T^{\mu \nu}(\rh) -T^{\mu \nu}(\sig) \right] \\
        & - (\partial_\mu \alpha) \left[ N^\mu (\rh) - N^\mu (\sig) \right] \\
        = & - \mathfrak{s}(\rho) + \nabla_\mu w^{\mu}(\sigma) \\
        & -  (\nabla _\mu \beta_\nu)  T^{\mu \nu}(\rh) - (\partial_\mu \alpha)  N^\mu (\rh)  \Big\}
        \leq 0.
    \end{split} \label{eq:localSecondLawLocalEquilibrium3}
\end{equation}
Here we wrote the change in relative entropy in a local way as was done already in eq.\ \eqref{eq:DeltaSrhosigmaLocal}.

\subsection{General exponential density matrices}\label{sec:BathGeneraExponential}
It is interesting to study a class of density matrices which one may call {\it general exponential} density matrices. This class generalizes global and local thermal equilibrium density matrices further and is of the form
\begin{equation}
    \sigma = \frac{1}{Z} \exp\left[ \int d\Sigma_\mu \left\{ - h^\mu_{\alpha\beta} T^{\alpha\beta} - l^\mu_{\alpha} N^\alpha \right\} \right],
    \label{eq:sigmaGeneralExponential}
\end{equation}
with the (non-equilibrium) partition function
\begin{equation}
    Z = \text{Tr} \left\{ \exp\left[ \int d\Sigma_\mu \left\{ - h^\mu_{\alpha\beta} T^{\alpha\beta} - l^\mu_{\alpha} N^\alpha \right\} \right] \right\}.
\end{equation}
The energy-momentum tensor $T^{\alpha\beta}(x)$ and particle number current $N^\alpha(x)$ should here be considered as operators, while the coefficients $h^\mu_{\alpha\beta}(x)$ and $l^\mu_{\alpha}(x)$ are parameter fields.

Note that for a given hypersurface $\Sigma$, with local normal vector $n_\mu(x)$, only some components of the parameter fields, namely the contractions $n_\mu(x) h^\mu_{\alpha\beta}(x)$ and $n_\mu(x) l^\mu_{\alpha}(x)$ actually enter eq.\  \eqref{eq:sigmaGeneralExponential}. In contrast, the components orthogonal to the surface normal vector field $n^\mu(x)$ could be changed without changing the density matrix $\sigma$. In this sense, there are precisely as many independent components of the parameter fields as there are components of the energy-momentum tensor and particle number current. One can understand the parameter fields as Lagrange multiplier fields that can realize unrestricted local expectation values $\langle T^{\mu\nu}(x) \rangle$ and $\langle N^\mu(x) \rangle$. In particular, these expectation values are not bounded to be of the thermal equilibrium or ideal fluid form.

A nice feature of eq.\ \eqref{eq:sigmaGeneralExponential} is that one can again express the von Neumann entropy $S(\sigma)$ in terms of expectation values and the partition function,
\begin{equation}
    \begin{split}
        S(\sigma) & = - \text{Tr}\left\{ \sigma \ln \sigma \right\} \\
        & = \ln(Z) + \int d \Sigma_\mu \left\{ h^\mu_{\alpha\beta} T^{\alpha\beta}(\sigma) + l^\mu_{\alpha} N^\alpha(\sigma) \right\}. 
    \end{split}
\end{equation}
If now the logarithm of the partition function (the Schwinger functional) is itself local and can be written as in eq.\ \eqref{eq:PartitionFunctionLocalEq} (this must be tested), one may define for the class of density matrices in \eqref{eq:sigmaGeneralExponential} the local entropy current
\begin{equation}
    s^\mu(\sigma) =  - h^\mu_{\alpha\beta} T^{\alpha\beta}(\sigma) - l^\mu_{\alpha}  N^\alpha(\sigma)+ w^\mu(\sigma),
    \label{eq:defEntropyCurrentSigma}
\end{equation}
such that
\begin{equation}
    S(\sigma) = - \int d\Sigma_\mu s^\mu(\sigma).
\end{equation}
The relation $\Delta S(\sigma) \geq 0$ implies again the local relation $\nabla_\mu s^\mu(\sigma) \geq 0$ for the class of states \eqref{eq:sigmaGeneralExponential}, provided their form is preserved by the corresponding CPTP time evolution map.

Another advantage of the exponential form \eqref{eq:sigmaGeneralExponential} is that one can determine the relative entropy of some state $\rho$ relative to such a state $\sigma$,
\begin{equation}
    \begin{split}
        & S(\rho \| \sigma)  = \text{Tr} \left\{ \rho \left( \ln \rho - \ln \sigma \right) \right\} \\
        & = - S(\rho) + \ln(Z_\sigma) + \int d\Sigma_\mu \left\{ h_{\alpha\beta}^\mu T^{\alpha\beta}(\rho) + l^\mu_\alpha N^\alpha(\rho) \right\} \\
        & = - S(\rho) + S(\sigma)  + \int d\Sigma_\mu \, \Big\{ h_{\alpha\beta}^\mu \left[T^{\alpha\beta}(\rho) - T^{\alpha\beta}(\sigma) \right] \\
        & \quad\quad + l^\mu_\alpha \left[ N^\alpha(\rho) - N^\alpha(\sigma) \right] \Big\} \\
        & = - S(\rho)   + \int d\Sigma_\mu \, \Big\{ - s^\mu(\sigma) + h_{\alpha\beta}^\mu \left[T^{\alpha\beta}(\rho) - T^{\alpha\beta}(\sigma) \right] \\
        & \quad\quad + l^\mu_\alpha \left[ N^\alpha(\rho) - N^\alpha(\sigma) \right] \Big\}.
    \end{split}\label{eq:relEntropyGeneralExpState}
\end{equation}
In the last equation we have used the definition of the entropy current $s^\mu(\sigma)$ associated with the density matrix $\sigma$ in eq.\  \eqref{eq:defEntropyCurrentSigma}.

Let us stress that \eqref{eq:sigmaGeneralExponential} for some given form of the parameter fields $h^\mu_{\alpha\beta}(x)$ and $l^\mu_\alpha(x)$ is in general not the result of a unitary time evolution. Instead, the evolution operators from one hypersurface  to another is a completely positive, trace-preserving (CPTP) map $\mathcal{N}$. In this sense, the class of density matrices in \eqref{eq:sigmaGeneralExponential} should be seen as describing open quantum systems. Applying the same evolution map to $\rho$ leads as before to an inequality as in eq.\ \eqref{eq:monotonicityRE}. This can again be made local, in generalization of but analogous to eq.\ \eqref{eq:localSecondLawLocalEquilibrium3}. 

\section{Conclusions and Outlook}
\label{sec:Conclusions}

We have investigated here how the second law of thermodynamics can be formulated with quantum relative entropy. For open quantum systems, which can exchange quantum information with an environment or heat bath, the time evolution is not unitary but given by a more general completely positive trace-preserving (CPTP) map. For classes of CPTP maps that leave equilibrium states invariant, the second law is a consequence of the monotonicity property of relative entropy. We have recalled this construction for generic quantum states in section \ref{sec:thermosecondlaw}.

Our main focus here was, however, to investigate local versions of the second law from this perspective, as they are being used for example in relativistic fluid dynamics. In a relativistic quantum field theory it is useful to consider besides global time evolution also evolution maps between more general Cauchy (hyper) surfaces, and this is particularly convenient to investigate local dynamics. We have specifically concentrated on situations where the Cauchy surfaces change only in a localized space-time region bounded by two light cones. For quantum field theories coupled to an external bath fluid, we have formulated local versions of the second law in terms of relative entropy. We have also discussed how the same relation could be formulated with relative {\it entanglement} entropy. The construction works with an external bath fluid in global equilibrium such that a reference state that is itself in equilibrium is left invariant, but it can also be extended to more general situations where this reference state is a local generalization of the equilibrium state or an even more general density matrix of exponential form in the energy-momentum tensor and conserved particle current operators.

In future work it would be important and interesting to check the scenarios we have laid out here for concrete and realistic quantum field theories. This implies also that appropriate (functional) methods need to be developed.

We have concentrated here on open quantum systems where the second law is in fact easier to understand than for closed quantum systems. For the latter, the evolution with time or between Cauchy surfaces is actually unitary, so that the global von Neumann entropy for the entire system is conserved. Also the relative entropy between two density matrices is then conserved.

Interestingly, a local version of the second law and local thermalization may nevertheless arise, as long as one considers only local observables. As we have discussed in section \ref{sec:LocalThermalEquilibrium}, it is possible (and in fact likely) that the quantum information spreads with time over space in the sense that further quantum entanglement between different spatial regions is generated. Even for a state that is far from global equilibrium, the reduced density matrix for some region may be equivalent to the reduced density matrix of a global thermal equilibrium state, for example, such that their relative entanglement entropy vanishes. Thermalization could occur locally but not globally. For a finite and isolated quantum system, a similar scenario would be in conflict with the possibility of quantum recurrences, but for a relativistic quantum field theory in an infinite space it is likely that these cannot occur.

In such a scenario, relativistic fluid dynamics would arise as an approximation to the full quantum field dynamics that describes local observables, but neglects some amount of non-local entanglement. It may be possible to approximate the originally unitary time evolution with a suitable coarse-grained variant constructed along these lines and this would then be a (C)PTP map similar to the one that arises for open quantum systems. In the future it would be interesting to investigate this scenario in more detail for concrete quantum field theories, both theoretically and -- within model systems -- experimentally.

\section*{Acknowledgements}
This work is supported by the Deutsche Forschungsgemeinschaft (DFG, German Research Foundation) under Germany's Excellence Strategy EXC 2181/1 - 390900948 (the Heidelberg STRUCTURES Excellence Cluster), SFB 1225 (ISOQUANT) as well as FL 736/3-1.


\bibliography{references.bib}

\end{document}